# Anisotropy Dependence of Irreversible Switching in Fe/SmCo and FeNi/FePt Exchange Spring Magnet Films

Joseph E. Davies,[a)] Olav Hellwig,[b), c)] Eric E. Fullerton,[b)] J. S. Jiang,[d)]
S. D. Bader,[d)] G. T. Zimányi[a)] and Kai Liu[a),*]

[a)] Department of Physics, University of California, Davis, CA 95616
[b)] Hitachi Global Storage Technologies, San Jose, CA 95120
[c)] BESSY GmbH, 12489 Berlin, Germany
[d)] Materials Science Division, Argonne National Laboratory, Argonne, IL 60439

## Abstract

Magnetization reversal in exchange-spring magnet films has been investigated by a First-Order Reversal Curve (FORC) technique and vector magnetometry. In Fe/epitaxial-SmCo films, the reversal proceeds by a reversible rotation of the Fe soft layer, followed by an irreversible switching of the SmCo hard layer. The switching fields are clearly manifested by separate steps in both longitudinal and transverse hysteresis loops, as well as sharp boundaries in the FORC distribution. In FeNi/polycrystalline-FePt films, particularly with thin FeNi, the switching fields are masked by the smooth and step-free major loop. However, the FORC diagram still displays a distinct onset of irreversible switching and transverse hysteresis loops exhibit a pair of peaks, whose amplitude is larger than the maximum possible contribution from the FeNi layer alone. This suggests that the FeNi and FePt layers reverse in a continuous process via a vertical spiral. The successive vs. continuous rotation of the soft/hard layer system is primarily due to the different crystal structure of the hard layer, which results in different anisotropies.

PACS numbers: 75.60.Jk, 75.30.Gw, 75.60.-d, 75.70.Cn

Exchange-spring magnets are an important class of artificially structured materials, initially proposed for permanent magnet applications.[1-3] Their magnetization reversal processes are fascinating[4-9] and important, such as for thermally assisted magnetic recording applications.[10] The basic structure consists of a magnetically hard / soft bilayer, which has been extended to multilayer[11] and nanocomposite[12] structures. Furthermore, the constituent compositions and growth parameters can be varied to tune the magnetic properties. For example, the switching fields of the soft and hard layers are measures of the interlayer coupling strength and the overall film anisotropy, crucial for applications. In epitaxial bilayer spring magnets, the switching fields of individual layers can be conventionally determined from steps in the major loops.[4] However, in polycrystalline samples, particularly with a thin soft layer, the major loop does not display any clear step and the determination of switching fields becomes difficult.[6, 7]

In this work, we examine the effect of the hard layer crystallinity and anisotropy on the magnetization reversal processes using a first order reversal curve (FORC) method and vector magnetometry. We show that even for films with step-free major loops the switching fields can be *quantitatively* determined from the FORC distribution, which provides direct access to irreversible switching processes. For Fe/epitaxial-SmCo samples, the Fe layer reverses its magnetization first via a reversible magnetization rotation, followed by an abrupt and irreversible SmCo switching. For FeNi/polycrystalline-FePt samples, the FeNi and FePt layers reverse simultaneously, by a predominantly irreversible magnetization rotation, forming a spiral. The chirality of such a spiral is preserved well beyond the apparent saturation of the hard layer.

Samples of Fe/SmCo and FeNi/FePt were grown by dc magnetron sputtering. For Fe/SmCo samples, MgO(110) substrates were used with an epitaxial 200-Å Cr(211) buffer layer. A 200-Å SmCo layer was then deposited at a substrate temperature of 600°C with a nominal composition of $Sm_2Co_7$, co-sputtered from elemental targets.[13] Finally an Fe layer was deposited at 300 - 400°C with thickness values in the range of 25 – 200 Å and capped with a 50-Å Cr layer. For FeNi/FePt samples, glass substrates with a 15-Å Pt seed layer were used. A 200-Å $Fe_{55}Pt_{45}$ layer was co-sputtered from elemental targets at 420°C. A $Ni_{80}Fe_{20}$ layer was then sputtered from an alloy target at 150°C with a thickness in the range of 50 - 800 Å, and finally capped with Pt.

Structural characterizations of the films have been carried out by x-ray diffraction. For Fe/SmCo films, the hard phase of SmCo is an epitaxial ($1\bar{1}00$) layer, with a strong in-plane uniaxial anisotropy along its *c*-axis. For FeNi/FePt films, the FePt hard layer is in the highly anisotropic $L1_0$ phase. It is polycrystalline with a (111) texture. Additional structural characteristics of the sample can be found in prior publications.[4, 6, 11, 13, 14]

Magnetic properties have been measured using an Alternating Gradient Magnetometer (AGM) and a vector coil Vibrating Sample Magnetometer (VSM) at room temperature. For the Fe/SmCo series the magnetic field is applied parallel to the in-plane magnetic easy axis of the SmCo hard layer, whereas for the FeNi/FePt series it is applied in-plane with an arbitrary orientation.

The AGM is used to measure a large number ($\sim 10^2$) of First-Order Reversal Curves (FORC's) in the following manner. After saturation, the magnetization *M* is measured starting from a reversal field $H_R$ back to positive saturation, tracing out a FORC. A family of FORC's is measured at different $H_R$, with equal field spacing, thus filling the interior of the major hysteresis loop (Figs. 1a & 1c). A FORC distribution is defined by a mixed second order derivative: $\rho(H_R, H) \equiv -\frac{1}{2}\frac{\partial^2 M(H_R, H)}{\partial H_R \partial H}$,[15-17] which eliminates the purely reversible components of the magnetization.[18] Thus any non-zero $\rho$ corresponds to *irreversible* switching processes.[17] For each FORC in Fig. 1a with a specific reversal field $H_R$, the magnetization $M$ is measured with increasing applied field $H$; the corresponding FORC distribution $\rho$ in Fig. 1b is represented by a horizontal line scan at that $H_R$ along $H$. For example, three line-scans corresponding to the reference points in Fig. 1a are represented by dashed lines in Fig. 1b. As $H_R$ decreases and the family of FORC's is measured, $\rho$ is scanned in a "top-down" fashion in the $H$-$H_R$ plane, mapping out the irreversible processes.

For the Fe/epitaxial-SmCo samples, the major loops clearly display *two* separate stages of reversal. For example, in Fe (100 Å)/SmCo (200 Å), a sudden decrease of magnetization around -2.5 kOe (reference point 1 in Fig. 1a) corresponds to the onset of the soft Fe layer reversal. At -7.8 kOe (point 2), a precipitous drop in magnetization indicates the sudden switching of the hard SmCo layer. Finally the reversal is completed at around -10 kOe (point 3). We notice that the FORC's nearly always overlap between points 1 & 2, but are well separated between points 2 & 3, confirming the corresponding reversible and irreversible switching within those field ranges.

In the resulting contour plot of the FORC distribution, for -2.5 < $H_R$ < -7.8 kOe (line 1 & 2 in Fig. 1b), there is no appreciable feature, consistent with the reversible switching of the soft Fe layer. However, there is a clear and sudden onset of FORC feature (where $\rho$ becomes non-zero) at around -7.8 kOe (line 2). This irreversibility onset coincides with the SmCo hard layer switching seen in the major loop (Fig. 1a, point 2).

Finally, beyond $H_R < -10$ kOe (line 3), the sample reaches negative saturation and any further field sweep would overlap and trace back up along the perimeter of the major loop. The FORC distribution then returns back to the $\rho = 0$ plane.[19]

For the FeNi/FePt samples, the major loops are distinctly different from the Fe/SmCo series since we observe at best *one* sharp magnetization drop during a field sweep, corresponding to the onset of reversal.[6] As the FeNi layer thickness decreases, this onset becomes more gradual as the soft layer couples more strongly onto the hard layer. For example, the major loop of a FeNi (100 Å)/FePt (200 Å) film no longer have any sudden magnetization drop (Fig. 1c). Over most of the reversal field range, $H_R < -1.2$ kOe (point 1 in Fig. 1c), the adjacent FORC's do not overlap. They fill the interior of the major loop rather evenly, indicating irreversible switching during the entire reversal. The corresponding FORC distribution shows only a single onset of irreversibility around $H_R = -1.2$ kOe (line 1 in Fig. 1d). This implies the hard and soft layers switch together, unlike the Fe/SmCo series where the soft layer reverses much earlier than the hard layer. Interestingly, the FORC distribution, and thus irreversible switching, persists for $H_R < -1.2$ kOe, even beyond $H_R < -10$ kOe where the major loop appear saturated.

The onset and endpoint of irreversible switching can be viewed more readily from the projection of the FORC distribution $\rho$ onto the $H_R$ axis. Such a projection is equivalent to integrating $\rho$ along $H$, leading to $\int \frac{\partial^2 M(H_R, H)}{\partial H_R \partial H} dH = \frac{dM(H_R)}{dH_R}$, which also characterizes the switching field distribution (SFD). Conventionally the SFD information is determined from the DC-Demagnetization (DCD) remanence curve,[20, 21] by taking the full width at half maximum (FWHM) of the $dM_r(H_R)/dH_R$ curve, where $M_r(H_R)$ is the zero field magnetization along a FORC with reversal field $H_R$. Thus the FORC projection and the DCD methods are similar.[22] For comparison, DCD remanence curves were extracted from the FORC data and their derivatives were calculated. The SFD determined from DCD agrees fairly well with that from the FORC projection method, as shown in Figs. 2c and 2d. However, there could be subtle differences (Fig. 2d) since the DCD method only references the remanent state, whereas the FORC method follows irreversible switching along the entire reversal curve up to positive saturation.

In order to distinguish magnetization rotation vs. domain wall nucleation and motion, we have used vector coil VSM to measure magnetization components parallel (longitudinal, $M_{//}$) and perpendicular (transverse, $M_\perp$) to the applied field during a field cycle. Representative $M_{//}$ and $M_\perp$ loops for Fe/SmCo are shown in Fig. 2a. Broad $M_\perp$ peaks (or steps) with large amplitudes have been observed, which can be attributed to unidirectional rotation of the moments during reversal. The onset of $M_\perp$ peak along each field sweep direction coincides with the initial drop of $M_{//}$, or the start of the soft layer rotation. The maximum $M_\perp$ is comparable to the saturation magnetization of the Fe soft layer (shown by dashed lines in Fig. 2a), indicating that most of the Fe moments have rotated prior to the abrupt switching of the hard layer. Furthermore, the $M_\perp$ peaks in Fe/SmCo samples occur on opposite sides of the longitudinal field (inverted relative to the origin) during descending and ascending-field sweeps. This is due to an imperfect alignment of the SmCo uniaxial easy axis with the applied field. The projection of the FORC diagram along $H_R$ (Fig. 2c, open squares) confirm that all of the irreversible switching occur during the second stage of switching (~ -8 kOe).

For the FeNi/polycrystalline-FePt, $M_\perp$ (Fig. 2b, open circles) shows sharp peaks, whose magnitudes are larger than the saturation magnetization of the FeNi soft layer alone (dashed lines). The extra transverse moment must come from the FePt hard layer, thus confirming the bilayer co-rotation. Also, these peaks now occur on the same side of the applied field during both descending and ascending field sweeps (mirror symmetry relative to H=0). It is interesting to note that such large $M_\perp$ peaks are observed when the hard layer has no well-defined in-plane anisotropy [the FePt (001) easy axis lies at an angle to the film normal with random in-plane distribution]. By systematically rotating the sample with respect to the in-plane applied field, we did observe a small residual anisotropy (absent in films of polycrystalline FePt alone), which differentiates the rotation direction. During a descending-field sweep, once the bilayer co-rotation starts, a spiral structure winds from FeNi into FePt (upward, as in Fig. 2b), forming a domain wall parallel to the interface, which is consistent with earlier studies.[8, 23] At -15kOe, even though the major loop appears saturated, the sample has not reached a true saturation and the chirality of the spiral is still preserved by some of the FePt grains, as evidenced by the persistent tail in the FORC distribution shown in Fig. 1d. In the subsequent increasing-field sweep, the spiral unwinds from the same direction (upward again, as in Fig. 2b), leading to a second $M_\perp$ peak that is mirror-symmetric to the first one. It is not until field cycling to ± 70 kOe, in a transverse SQUID magnetometer, are we finally able to suppress the $M_\perp$ peaks and eliminate the residual anisotropy. The interesting rotation process and the preservation of the chirality are important for applications of spring magnets, such as in thermally assisted recording.[10]

In conclusion, we have *quantitatively* determined the switching fields of Fe/SmCo and FeNi/FePt and the effect of hard layer crystallinity using a FORC method and vector magnetometry. In epitaxial Fe-SmCo films with well defined in-plane uniaxial anisotropy of SmCo, magnetization reversal is initiated by a reversible rotation of the Fe soft layer, followed by an abrupt and irreversible switching of the SmCo hard layer. The rotation of the Fe soft layer is inversion-symmetric relative to zero field. In FeNi/polycrystalline-FePt films with random in-plane anisotropy of FePt, magnetization reversal is predominantly by irreversible continuous rotation of the bilayer. The rotation of the bilayer is mirror-symmetric relative to zero field as the domain wall chirality is preserved in the random anisotropy hard layer well beyond the apparent saturation.

This work has been supported by NSF (EAR-0216346), DOE (BES-MS contract #W-31-109-ENG-38), and University of California (CLE). We thank C. P. Pike, K. L. Verosub, R. T. Scalettar, G. Acton, A. Roth, and A. Berger for helpful discussions.

**Figure Captions**

Fig. 1: (Color online and in print) Families of first order reversal curves for films of (a) Fe (100Å)/SmCo (200Å) and (c) FeNi (100 Å)/FePt (200Å) film, where the first point of each reversal curve is shown by a black dot. Contour plots of the corresponding FORC distribution are shown in (b) and (d) respectively, versus applied field $H$ and reversal field $H_R$. Reference points are marked in (a, b) and (c, d) to illustrate the different reversal stages.

Fig. 2: Longitudinal (solid circles) and transverse (open circles) hysteresis loops of (a) Fe (100Å)/SmCo (200Å), and (b) FeNi (100Å)/FePt (200Å). Saturation magnetization of the soft layers is marked by the dashed lines. (c) & (d), the projection of $\rho$ along $H_R$ (open squares) and DCD remenance curve (closed squares) are shown to illustrate the onset of irreversible switching during the decreasing-field sweep.

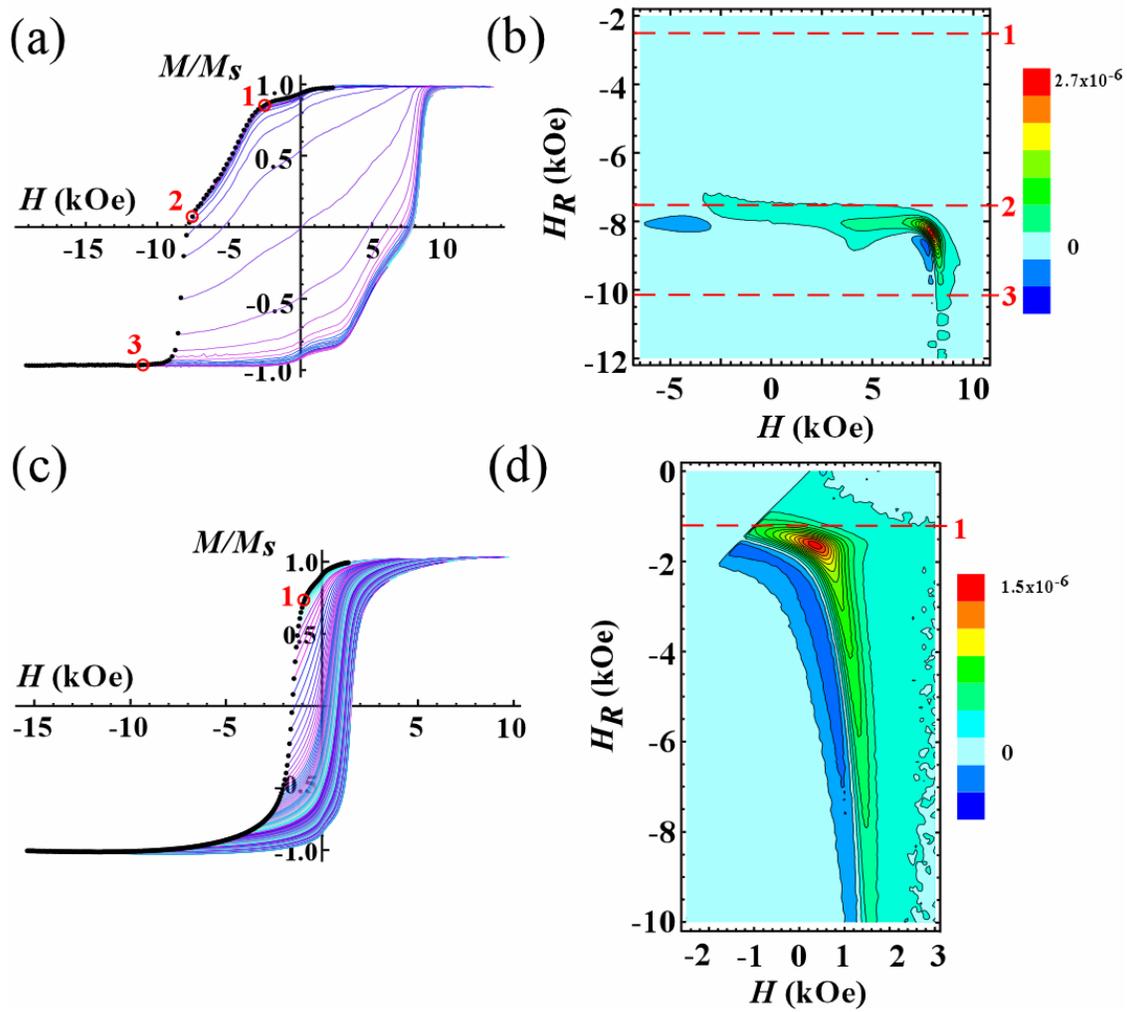

Fig. 1

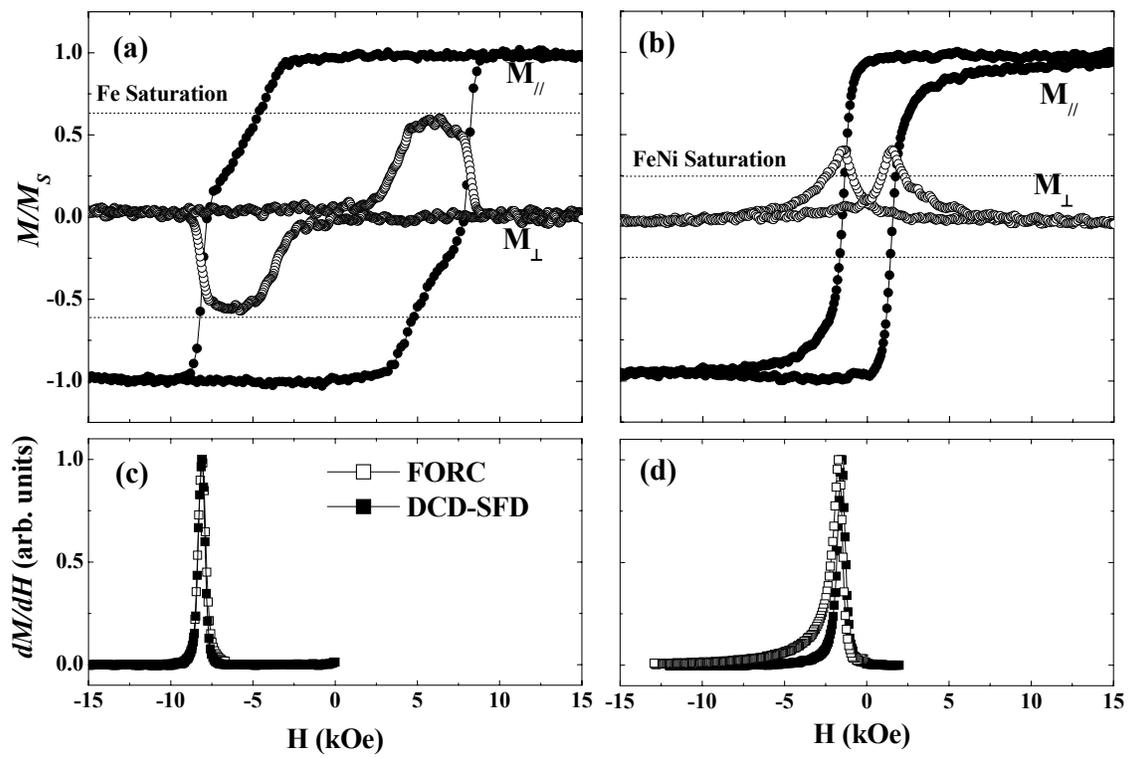

**Fig. 2**